\documentclass{aastex}
\usepackage{emulateapj5}
\def\gtrsim{\mathrel{\hbox{\rlap{\hbox{\lower4pt\hbox{$\sim$}}}\hbox{$>$}}}}
\let\ga=\gtrsim
\def\lesssim{\mathrel{\hbox{\rlap{\hbox{\lower4pt\hbox{$\sim$}}}\hbox{$<$}}}}
\let\la=\lesssim
                                \begin{document}

                                \title{
Detection of Bulk Motions in the ICM of the Centaurus Cluster
                                }

                                \author{
Renato A. Dupke \& Joel N. Bregman
                                }

                                 \affil{
University of Michigan, Ann Arbor, MI 48109-1090
                                }

                                \begin{abstract}
Several recent numerical simulations of off-center cluster mergers predict that significant 
angular momentum with associated velocities of
 a few $\times$10$^{3}$ km~s$^{-1}$ can be imparted to the resulting cluster. Such gas bulk velocities can be detected by the 
 Doppler shift of X-ray spectral lines with {\sl ASCA} spectrometers. Using two {\sl ASCA} 
observations of the Centaurus cluster (Abell 3526), we produced a velocity map for the gas in
the cluster's central regions.
We also detected radial and azimuthal gradients in temperature and metal abundance distributions, 
which seem to be associated with the infalling sub-group centered  
at NGC 4709 (Cen 45). 
More importantly, we found a significant 
($>$99.8\% confidence level) velocity gradient along a line near-perpendicular to the direction of 
the incoming sub-group and with a maximum velocity 
difference of $\sim$ 3.4$\pm$1.1 $\times$10$^{3}$ km s$^{-1}$. It is unlikely 
(P $<$ 0.002) that the observed velocity
gradient is generated by gain fluctuations across the detectors.
While the observed azimuthal temperature and abundance variations can be attributed to the interaction
with Cen 45, we argue that the intracluster gas velocity gradient is more likely due to a previous
off-center merging event in the main body of the Centaurus cluster.

                                \end{abstract}

				
~~~~~~~~~~~~~~~~~~~~~~~~~~~~~~~~~~~~~~~~~~~~~~~~~~~~~~~~~~~~~~ApJ in press

                                \section{
Introduction
                                }
				
Most of the visible mass in clusters is in the form of a very hot (T $\sim$ 10$^{7-8}$ K) diffuse 
X-ray emitting plasma enriched with heavy elements that permeates the cluster, the so-called intracluster 
medium (ICM). 
Galaxy clusters are believed to be formed by the infall/merging of smaller scale 
systems (bottom-up hierarchical scenario). The merging of sub-clumps induces 
several long lasting intracluster gas signatures that can be observed in the X-ray range, such as
temperature and density inhomogeneities, destruction of cooling flows and metal abundance 
gradients and generation of gas bulk velocities.
Spatial distributions of surface brightness and gas temperature
can be directly compared to numerical hydrodynamical+N body simulations of 
cluster mergers to provide a picture of the evolutionary stage of clusters (e.g. Evrard 1990; Katz \& White 1993; 
Roettiger, Burns \& Loken 1993,1996; Schindler \& Muller 1993; 
Pearce, Thomas \& Couchman 1994; Navarro, Frenk \& White 1995; Evrard, Metzler, \& Navarro 1996; 
Roettiger, Loken \& Burns 1997, Ricker 1998; Takizawa \& Mineshige 1998; Burns et al. 1999; Takizawa 1999, 
2000 and references therein).

{\sl ASCA} observations have shown significant gas temperature variations in many clusters, which are  
often attributed to recent mergers and this suggests that galaxy clusters are still forming. However, 
when determining the precise evolutionary stage of the intracluster gas, the distribution of gas temperature 
and surface brightness alone have limitations. This is because they represent only a 2-D projection of a 3-D system. 
In order to reduce degeneracies and 
to determine the cluster evolution stage more precisely it is extremely important to obtain as much information 
as possible about the dynamics of the intracluster gas. Several recent numerical 
simulations of off-center cluster mergers predict that significant 
angular momentum can be imparted to the resulting cluster with velocities approaching 
 a few thousand km~s$^{-1}$. Simulations also show  that gas bulk flows may last for several Gyr after the 
 merger (e.g. Roettiger, Loken \& Burns 1997, Ricker 1998; Roettiger, Stone, \& Mushotzky 1998; 
 Takizawa \& Mineshige 1998; Burns et al. 1999; Takizawa 1999, 2000; Roettiger \& Flores 2000). 
Therefore, measurements of radial velocity distribution of the intracluster gas provides a very important 
additional source of information that, coupled with temperature and surface brightness distributions, 
can be  directly compared to numerical simulations to determine the evolutionary stage of galaxy clusters. 
 
Radial velocity measurements of the intracluster gas can be obtained from X-ray observations, 
but require accurate determinations of spectral line centroids. The precision with which a line centroid 
can be measured is a function 
of the square root of the number of line photons, the FWHM of the instrument, the line energy and,
in practice, it strongly depends on knowledge of the instrumental gain (conversion between pulse height and energy).
The instruments on-board {\sl ASCA} have the capability of measuring directly such ICM bulk 
velocities. Furthermore, {\sl ASCA}'s spectrometer's gain calibration is already well-understood 
as to allow reliable velocity measurements.

Recently, using the GISs on-board {\sl ASCA}, we have detected a velocity gradient in the intracluster 
gas of the Perseus cluster. The spectral analysis of 8 different {\sl ASCA} pointings encompassing a region 
of $\sim$ 1 h$_{50}^{-1}$ Mpc radius around the center of the Perseus cluster (Dupke \& Bregman 2000, 2001), 
showed a significant velocity gradient of $\ge$ 1000 km s$^{-1}$Mpc$^{-1}$ at the 90\% confidence.
However, the GISs have worse energy resolution than the SISs, which severely limited the ability to 
precisely constraint the velocity gradient amplitude. Ideally, one would like to use the higher 
energy resolution of the SISs to measure ICM bulk velocities.

The Centaurus cluster provides us with an excellent opportunity to measure intracluster gas velocities more precisely. 
It is a bright, nearby cluster, with high metal abundances (well isolated spectral lines) 
and shows signs of interaction with an infalling subgroup (Cen 45) located $\sim$ 13-18$^\prime$ from the center, which 
is associated with the galaxy NGC 4709. Furthermore, it has been an {\sl ASCA} target for a
long exposure ($\sim$ 90 ksec) taken in 1-CCD mode with the best calibrated CCDs. These observational 
characteristics make Centaurus an optimal target for gas velocity measurements.

In this paper we analyze the central regions of Centaurus and determine 
the velocity structure of this system using data from two different {\sl ASCA} pointings. 
In sections 2 \& 3 we describe
the general characteristics of the Centaurus cluster, of the {\sl ASCA} pointings analyzed and 
of data reduction procedures. In section 4 we investigate whether the velocity differences between Cen 30 and 45
previously observed optically by other authors can be corroborated by X-ray velocity analysis using the GISs. 
In section 5 we present the spectral fitting results of SIS data for the observation with the longest exposure (taken in 1995)
and estimate the effects of gain variations in our results as a function of frequency and position.
In section 6 we show the results of the spectral analysis of another {\sl ASCA} pointing taken 2 years
earlier, with a shorter exposure. In section 7 we briefly discuss the nature of the small and large scale bulk 
motions detected in Centaurus.


\section{The Centaurus Cluster}

Centaurus (Abell 3526) is classified as Bautz-Morgan type I, it has an optical redshift 
of 0.0104 and is one of the closest X-ray bright clusters of galaxies. It has a small cooling flow 
and the estimated accretion mass rate is $<$ 
30-50 solar masses per year (e.g. Matilsky, Jones \& Forman 1985; Allen \& Fabian 1994; White, Jones \&
Forman 1997; Peres et al. 1998). Its X-ray emission is relatively smooth, slightly elliptical in
shape and is strongly peaked on the cD galaxy NGC4696. The average temperature of the 
gas in the cluster is $\sim$ 3.5 keV (e.g. Mitchell \& Mushotzky 1980; Matilsky et al. 1985; 
Thomas, Fabian \& Nulsen 1987; Edge \& Steward 1991). {\sl ASCA} GIS \& SIS 
analyses of the central region of Centaurus have 
shown evidence of a strong central metal abundance enhancement (Allen \& Fabian 1994; Fukazawa et al. 1994)
varying from supersolar in the very 
central regions down to 0.3 solar at $\sim$ 13$^\prime$ (1$^\prime$ at Centaurus distance is equivalent
to $\sim$ 19 h$^{-1}_{50}$kpc). Furthermore, the region where metal abundances are enhanced is dominated by SN Ia 
ejecta (Allen et al. 2000) indicating the presence of a radial ``chemical gradient''. 

Early studies of galaxy velocities in the Centaurus cluster have shown the existence of
bimodality in velocity space (Lucey, Currie \& Dickens 1986a,b). Two galaxy groups are clearly separated:
The main group (Cen 30) which is centered on the cD galaxy (NGC 4696) shows an average radial velocity of 
$\sim$3000 km~s$^{-1}$ and a velocity dispersion of $\sim$590 km~s$^{-1}$. The second group (Cen 45) is
associated with the galaxy NGC 4709 at $\ga$15$^\prime$ from NGC 4696. It has an average radial velocity
of $\sim$ 4600 km~s$^{-1}$ and a velocity dispersion of $\sim$ 280 km~s$^{-1}$. Lucey et al. (1986a,b) 
concluded that the Cen 45 and Cen 30 are at the same distances and that Cen 45 is being accreted 
by the Centaurus cluster, based on a detailed 
analysis of the color-magnitude, luminosity function and galaxy radius distribution. Improved redshift measurements
of the central regions of Centaurus confirmed the galaxy velocity bimodal distribution: the ``true'' cluster Cen 30
having an average velocity of 3397$\pm$139 km~s$^{-1}$ and a velocity dispersion of 933$\pm$118 km~s$^{-1}$, and 
Cen 45 having an average velocity of 4746$\pm$43 km~s$^{-1}$ and a velocity dispersion of 131$\pm4$3 km~s$^{-1}$ (Stein et al. 1997).

The X-ray surface brightness contours of Centaurus are slightly elliptical and NGC 4709 is located near the 
major axis (Allen \& Fabian 1994), suggesting that the elongation towards Cen 45 can be associated
with the elongation of the gravitational well along the collision axis. More recently, Churazov et al. (1999) 
has derived an X-ray temperature map 
of the Centaurus cluster using the {\sl ASCA} satellite and found a region where the intracluster gas is significantly 
hotter ($\ga$5 keV) than the surroundings ($\sim$3.5-4.0 keV). The region with higher gas temperature is apparently 
associated with Cen 45 and indicates the presence of shocks due to the infall of subcluster Cen45.

                                \section{
Data Reduction \& Analysis
                                }

{\sl ASCA} carries four large-area X-ray telescopes, each with its own detector: two Gas 
Imaging Spectrometers (GIS) and two Solid-State Imaging Spectrometers (SIS). Each 
GIS has a $50'$ diameter circular field of view and covers a energy range of 0.8--10 
keV, while each SIS has a $22'$ square field of view and covers an energy range of 0.4--10 keV.
For all pointings analyzed in this work we selected data taken with high and medium bit rates, 
with cosmic ray rigidity
values $\ge$ 6 GeV/c, with elevation angles from the bright Earth of $\ge20^{\circ}$, 
and from the Earth's limb of $\ge5^{\circ}$ (GIS) or $\ge10^{\circ}$ (SIS); we also excluded times when the 
satellite was affected by the South Atlantic Anomaly. Rise time rejection of particle 
events was performed on GIS data and hot and flickering pixels were removed from SIS data. 
We estimated the background from blank sky files provided by the {\sl ASCA} 
Guest Observer Facility. In the spectral fittings we used {\tt XSPEC} v10.0  
software (Arnaud 1996) to fit the GIS \& SIS spectra separately and simultaneously. The 
spectra were fitted 
using the {\tt mekal} and {\tt Vmekal} thermal emission
models, which are based on the emissivity calculations of  Mewe \& Kaastra (Mewe,
Gronenschild \& van den Oord  1985; Mewe, Lemen \& van den Oord 1986; Kaastra  1992),
with Fe L calculations by Liedahl et al. (1995). Abundances are 
measured relative to the solar photospheric values (Anders \& Grevesse 1989), 
in which Fe/H=$4.68\times10^{-5}$ by number. Galactic photoelectric absorption was 
incorporated using the {\tt wabs} model (Morrison \& McCammon 1983); Throughout the paper 
the hydrogen column density is assumed to be fixed to the Galactic value, i.e., 
8.07$\times$10$^{20}$cm$^{-2}$ (Dickey \& Lockman 1990 - W3nH tool\footnote{http://heasarc.gsfc.nasa.gov/cgi-bin/Tools/w3nh/w3nh.pl}), 
unless stated otherwise. Spectral channels were grouped to have at least 25 counts/channel. Since 1994, 
there have been 
increasing discrepancies between SIS and GIS in the low energy end of the spectra , due to 
a decrease in X-ray efficiency in the SISs\footnote{heasarc.gsfc.nasa.gov/docs/asca/watchout.html, 
also Hwang et al. 1999}; Therefore, energy ranges for both instruments were restricted to 
0.8--10 keV, in all spectral fittings. 

The central regions of the Centaurus cluster were observed by {\sl ASCA} in June of 1993 
(P93) and on July of 1995 (P95) for $\sim$ 24 ksec and 87 ksec, respectively; the characteristics 
of these two pointings are shown in Table 1. The original energy resolution of the
spectrometers on-board {\sl ASCA} were 8\% and 2\% at 5.9 keV for GISs 
and SISs, respectively. The SISs energy resolution has been steadily degrading with time 
(e.g. Dotani et al. 1997) due to the increases in dark current and decline in X-ray 
signal charge transfer efficiency (Rasmussen et al. 1998). For the best 
pointing analyzed in this work (P95) the energy resolution is  $\sim$  2.9\% and 3.1\% at 
6.0 keV for SIS 0 \& 1, respectively. Furthermore, for P95 the SIS observations were taken 
in 1-CCD mode using the best 
calibrated chips of each spectrometer (S0C1 and S1C3).

Since there is a mild cooling flow at the center of Centaurus we added a cooling 
flow component (Mushotzky \& Szymkowiak 1988) to the {\tt mekal} thermal emission 
model in the central region, 
to compare the temperature of the hot component in the central region with that of the outer
regions. We tied the maximum temperature of the cooling flow to the 
temperature of the thermal component, and we fixed the minimum temperature at 0.1 keV. 
The abundances of the two spectral components ({\tt mekal} and {\tt cflow}) were tied
together. We also applied a single (but variable) absorption to both spectral 
components. 

We have noticed an unusual gain variation in the SISs when comparing spectral fittings including the 
FeK complex to those excluding it. It is probably due to the assumption that the charge transfer 
inefficiency is independent of X-ray energy during the initial gain standard correction procedure (see below).
 Typically, spectral fittings that excluded the FeK line complex
showed higher best-fit redshift values. Since we are looking for relative 
velocity measurements this effect should not 
affect our analysis. However, we also show the results from spectral fittings excluding 
the FeK complex, where the energy range considered goes from 0.8-6.0 keV and are denoted by a sub-index FeL.

For P95 we noticed significant discrepancies of best-fit redshifts for some regions (e.g. P8) between 
SIS 0 \& 1 spectral fits. This is probably due to different spatial distribution of 
gain fluctuations in different chips. 
Therefore, we avoided performing simultaneous spectral fittings using these two spectrometers. 
The statistics are sufficiently good that radial 
velocity distributions can be measured precisely on each instrument separately. For P93 however, 
the statistics are not 
good enough to notice significant changes of the best-fit parameters 
when spectra from SIS 0 \& 1 are fitted separately. 
Therefore, for that pointing, we show only the results of joint spectral fittings. 

 For the analysis of the cluster's central regions with the SISs we selected a set of 9 different extraction regions 
 within 8$^\prime$ of the cluster's center using the SISs. The very central region included 
 the cluster core (P0), and the other regions were distributed
symmetrically around P0, centered 5$^\prime$ away and with a radius of
3$^\prime$. These regions are illustrated in Figure 1, where they are overlaid on
the X-ray surface brightness contours. This set of extraction regions was applied to 
both pointings (P95 \& P93). We applied a correction for the position 
angles between these pointings (19.6$^\circ$), so that each extraction region corresponded 
to the same position in the sky for both observations. 

Since P93 was taken in 4-CCD mode and the bulk of the cluster
emission was extended over 2 CCDs in both SIS 0 \& 1, we have taken into account 
the overall interchip gain differences  when comparing spectra from different CCDs (see below). 
In doing so, if an extraction region crossed over 2 CCDs we considered only the 
fraction of that region that included the largest number of counts in one CCD. This procedure 
minimizes the errors involved 
in the process of gain correction. Therefore, when comparing the results from pointings P93 
to P95 one should 
notice that the centroids of the extraction regions may be slightly ($<$ 1$^{\prime}$) shifted with respect to 
one another for P93 regions: P1 (shifted towards P2), P4 (shifted towards P3), P5 (shifted towards P6) and P8 
(shifted towards P7).

For the GIS analysis of the outer ($>$8$^\prime$) parts of Centaurus, we selected two sets of extraction
regions: in the first set, the region encompassing Cen 45 (circle centered 11$^\prime$ from NGC 4696 towards 
Cen 45 with a 8$^\prime$ radius) was compared to the average. The second set of regions 
was chosen analogously to that for the SIS analysis of the central 8$^\prime$ and it included
a central region (0-5$^\prime$) and 8 large regions 
symmetrically distributed around the center, at a distance of 12$^\prime$ from the center and with a radius 
of 5$^\prime$. Since the GISs have a poorer energy resolution than the SISs, we joined the 8 outer regions 
by adjacent pairs (1-2, 3-4, 5-6, 7-8) in order to improve photon 
statistics. These regions are also illustrated 
in Figure 1. The objective of the 
analysis of the outer regions was to check for corroboration in X-ray 
frequencies of the velocity difference seen in the optical analyses between Cen 45 and Cen 30. 
\section{GIS Global Velocity Measurements of Cen 45
                                        } 
As we mentioned previously, the galaxy velocity distribution in Centaurus is bimodal 
(Lucey, Currie \& Dickens 1986a,b). 
One of the velocity components is associated with 
the main body of the cluster (Cen30) and has a radial systemic velocity of $\sim$ 3000 km~s$^{-1}$ 
and another component
associated with the galaxy NGC 4709 $\sim$ 15$^\prime$ away from Cen 30 has a radial systemic velocity of $\sim$ 
4500 km~s$^{-1}$ and is generally denoted by Cen 45. 
We examined whether the optical velocity differences have an 
X-ray counterpart. We applied a similar technique that was used in the analysis
of velocity measurements in the Perseus cluster with the GISs on-board {\sl ASCA} 
(see Dupke \& Bregman 2001 for details).

The most obvious spatial configuration to test for velocity differences between the region associated
with Cen45 and Cen 30 is to extract spectra from a circular region large enough to encompass 
the bulk of Cen 45 (11$^\prime$ away from the center with a 8$^\prime$ radius -- denoted C45)
and compare the best-fit 
redshift values to the average (total field-of-view 
minus C45). The best-fit parameters of the spectral fittings of the C45 region and of the average 
region are shown in Table 2. C45 region is significantly hotter (T=5.47$\pm$0.27 keV) than the average
gas temperature of Centaurus (3.44$\pm$0.03 keV), and has a lower metal abundance 
(0.5$\pm$0.09 Solar) than the average (0.77$\pm$0.03 Solar), consistent with a central negative
radial abundance gradient. The best-fit redshift value for C45 is
found to be 3.27$^{+1.58}_{-1.13}$ $\times$ 10$^{-2}$, which is significantly ($>$90\% confidence limits) higher 
than the average redshift of the rest of the cluster 1.81$^{+0.05}_{-0.01}$ $\times$ 10$^{-2}$. 

The errors cited above
are spectral fitting errors and do not include any uncertainties related to gain variations across the detectors.
In order to estimate the latter we used a similar procedure to that used previously in the GIS velocity 
analysis of Perseus
(Dupke \& Bregman 2001). 
We used the GIS gain map determined from the Cu-K line at 8.048 keV 
(Idesawa et al. 1995) to estimate the velocity differences expected purely from residual gain variations 
across the detector. We compared the average gain values of the C45 region for each instrument with
that of the field-of-view excluding C45. 
The gain variations across each detector
are corrected in-house with the ftool ASCALIN V.9.0t. If we assume that there are still significant gain 
fluctuations across the detector and, conservatively, that these
fluctuations are on the same order as those of the pre-processed data then the expected difference 
between the two regions due 
purely to gain variations is $\sim$1870 km~s$^{-1}$ (C45 having the higher radial velocity). Therefore, the ``real'' 
velocity difference (not directly associated with gain variations) is 
$\sim$ 2.6$\pm$2.0 $\times$ 10$^{3}$ km~s$^{-1}$ (1-$\sigma$ errors), 
which is consistent with the optical velocity measurements of $\sim$ 1500 km~s$^{-1}$.

The gain scatter across the C45 region in GIS 2 \& 3 is relatively high 
(on the same order of the velocities that we are trying to measure) lowering the 
significance of the observed velocity differences. Therefore, we looked
for another spatial extraction configuration that minimized the gain fluctuations 
inside the extraction regions but
still encompassing Cen 45 and with enough
photon statistics to measure velocities correspondent to the optical measurements. The resulting 
configuration chosen is also shown in Figure 1 and it consists of 
4 regions symmetrically distributed around the central region 
(circle centered on the X-ray peak with a 5$^\prime$ radius). The four regions are pairs of adjacent circular regions 
(12$^\prime$ away with 5$^\prime$ radius). They are labeled as J12, J34, J56 \& J78. This configuration shows a 
reduction of intrachip gain scatter by $\sim$ 17\% for GIS 2 and $\sim$32\% for GIS 3
with respect to the previous configuration. The
high temperature region detected by Churazov et al. (1999) associated with Cen 45
is located mostly within J56. The best-fitting parameter values for this configuration are also shown in Table 2. It
can be seen that the only region which shows significantly different redshifts ($>$90\% confidence) with respect to the 
central region is J56 an the corresponding velocity difference is 
$\sim$ 1.77$\pm$0.96 $\times$ 10$^{3}$ km~s$^{-1}$ (1-$\sigma$ errors). 

The results above are consistent
with the optical measurements and indicate that significant velocity gradients in clusters can be 
detected even with the GISs, which have worse energy resolution than the SISs on-board {\sl ASCA}.
In the next sections we use the SISs to analyze the very central regions of the Centaurus 
cluster. Besides having a better energy resolution (by a factor of $\gtrsim$2.5), the SIS spatial gain 
fluctuations are more stable than that of the GIS. Therefore, radial velocities can be measured with 
a substantially higher precision.
                                        \section{
SIS Analysis of the Central Regions of Centaurus
                                         }
 \subsection{SIS Observation of July 1995 (P95)
                                        }
					
P95 is the best pointing analyzed in this work and probably one of the best pointings 
in the {\sl ASCA} archive for the analysis of gas bulk motions. The combination of long-exposure 
(Table 1), clock mode (1-CCD), and cluster's center positioning in the CCDs (roughly at the center of 
S0C1 and S1C3) makes this pointing near-optimal for 
velocity studies with {\sl ASCA}. 
The best-fit values for temperature, abundance and redshift obtained from spectral fittings of the 
central regions are shown in Table 3 
and plotted in Figures 2a,b as a function of the position angle for SIS 0 \& 1, respectively.
The indicated errors in Figures 2a,b are 90\% confidence limits. 

It can be seen from 
Figures 2a,b that the temperature distribution is consistent with the presence of a cooling flow 
(with a best fit mass deposition rate $\la$ 
10 solar masses per year). The best-fit temperatures are found to be typically slightly higher using 
SIS 1 data than those obtained with SIS 0. The temperature in the center is $\sim$2.68$\pm$0.05 keV
 and shows a positive gradient outwards up to a maximum value of 
$\sim$3.59$\pm$0.09 keV and $\sim$3.85$\pm$0.12 keV for SIS 0 \&1, respectively. The 90\% 
confidence limits for the best-fit values in the central region are indicated
by the two horizontal lines in Figures 2a,b. Furthermore, the observed temperature gradient
is not purely radial. One can notice a significant azimuthal temperature gradient, which is 
roughly symmetric and with a maximum towards the direction of Cen 45, which is the region
in between the two vertical lines in all plots. The azimuthal 
temperature distribution has only one axis of symmetry and, therefore,
is not consistent with the ``cross'' pattern shape of {\sl ASCA} PSF scattering.

The metal abundance distribution is consistent with the existence of a central abundance 
gradient (Figure 2a,b and Table 3). In general, if different elemental metal abundances were 
tied such as to produce solar abundance ratios the fits were 
significantly worse (often with $\chi^{2}_{\nu}$ $\approx$ 1.3). 
There was a significant improvement in the spectral fittings when individual metal 
abundances were let free to vary ($\chi^{2}_{\nu}$ $\approx$ 1), resulting in strongly non-solar 
abundance ratios. This is consistent
with the observed dominance of SN Ia ejecta in the central regions of this cluster 
(Allen et al. 2000). Furthermore, letting the individual elemental abundances free to vary
did not cause significant changes in the best-fit values of temperature, Fe abundance or redshifts. 
Therefore, we report here only the spectral fittings for an absorbed (Galactic) {\tt Vmekal} 
(abundances free to vary) spectral model 
(with an extra cooling flow component for the central region). The metal abundance within the inner 
3$\arcmin$ was found to be 1.13$\pm$0.05 solar and 1.23$\pm$0.06 solar, for the SIS 0 \& 1, 
respectively. The 90\% confidence limits for the abundance in the central region are shown 
by the horizontal lines in Figures 2a,b. 
One can see also indications of an azimuthal metal abundance gradient. In the outer 5$\arcmin$ the 
metal abundance 
drops significantly towards the direction of Cen 45 (P3, P4, P5, P6 \& P7 for SIS 0 and 
P4, P5 \& P7 for SIS 1).
The minimum metal abundance values were reached around pointings P4 \& P5 in both 
detectors, where the best-fit value is $\sim$0.86$\pm$0.1 solar. 

The azimuthal distributions of redshifts are shown in Figures 2a,b (bottom plot) as
a function of the position angle. The original SIS 0 best-fit redshift values 
were too close to 0 so that full determination of 90\% confidence errors was compromised in {\tt XSPEC}. 
Therefore, all redshifts in that chip were boosted up by a slope correction of 
0.9\%\footnote{where the new line energy is related to the old energy in the standard way,
i.e., $E_{new} = \frac{E_{old}}{slope} - intercept$}. 
This redshift-boosting does not affect our results since we are not 
interested in the absolute redshift values, but only in the redshift differences between regions. 
SIS 1 redshifts are displayed directly (without any gain boosting). 

From Figure 2a we can see that the SIS 0 data shows a very significant velocity gradient. 
Regions P3 \& P4 have significantly
lower velocities than regions P6, P7 \& P8 by $\sim$2.67$\pm$0.21 $\times$ 10$^{3}$ km~s$^{-1}$
for SIS 0 
and 2.73$\pm$0.27 $\times$ 10$^{3}$ km~s$^{-1}$ for SIS 1 (1-$\sigma$ errors). Both velocity-discrepant sets of 
regions have significantly different velocities (on the order of 
$\sim$1350 km~s$^{-1}$ at $>$ 90\% confidence level) with respect to the clusters center.
Although the velocity gradient is also seen in SIS 1 data, its significance is lower than that 
obtained using SIS 0 data. 
This is due to the significantly smaller effective number of counts of SIS 1 with respect to SIS 0 
(30\% less counts) and to the larger gain spatial variability of SIS 1.  
Figure 2b shows the azimuthal distribution of best-fit redshifts obtained using SIS 1 data. 
It can be seen that for the ``low velocity'' regions 
(P3 \& P4) the redshifts are significantly different from that of the central region and 
also from most of the regions in the ``high velocity'' hemisphere (P6 \& P8). 
There is one region (P8) for which analysis of SIS 0 \& 1 spectra
gives discrepant results. Although the distribution of radial velocities determined using SIS 1 are
less constrained than that determined using SIS 0, the overall azimuthal distributions of velocities
are consistent 
with one another. {\it The velocity gradient found with SIS 1 is maximum along the same direction 
(P4--P6) as that determined from SIS 0 and has the same magnitude 
($\sim$ 380 km~s$^{-1}$arcmin$^{-1}$)}. 
The velocity differences are unlikely caused by spectral modeling effects since regions with
discrepant velocities have similar temperatures and metal abundances.

To assess the significance of the velocity differences described in the previous paragraph, 
we chose two regions where we consistently found the maximum 
velocity differences, P4 and P6, and we applied the F-test. We simultaneously fitted 
 spectra from SIS 0 \& 1 for each of the two regions and analyzed $\chi^2$ variations associated to 
the change in the number of degrees of freedom. 
We compared the $\chi^2$ of fits which assumed the redshifts to be the same in the two 
projected spatial regions P4 \& P6, i.e.,
$z_{S0_{P4}} = z_{S1_{P4}} = z_{S0_{P6}} = z_{S1_{P6}}$,
to that of fits which allowed the redshifts in the two regions to vary independently, reducing 
the number of degrees of freedom by one. 
For the latter we tied the redshifts 
within each of the two regions together, i.e., 
$(z_{S0_{P4}} = z_{S1_{P4}}) \ne (z_{S0_{P6}} = z_{S1_{P6}})$, where $z_{Si_{Pj}}$ is
the redshift of region $j$ with instrument SIS $i$. This is justified by the lack of
significant changes in the best-fit redshifts obtained by different
instruments for the same region.
The difference between the $\chi^2$ of these two fits must follow a $\chi^2$ 
distribution with one degree of freedom (e.g. Bevington 1969). 
The value $F(n)=\frac{\chi^{2}(n-1)-\chi^{2}(n)}{\chi^{2}_{\nu}(n)}$, where $n$ is the number of 
fitting terms, is related to the probability that the 
additional term significantly contributes to the reduction of the $\chi^{2}$ statistic.
Comparison of the $F(n)$ from the fits with locked and unlocked redshifts to tabulated 
values of F indicates 
that the best-fit redshifts obtained for P4 \& P6 are discrepant at $>99.99$\% level. 
We show in Figure 3 the 90\% and 99\% confidence contours for two interesting parameters 
(redshifts) of regions P4 and P6 as well as the line correspondent to equal redshifts.

 \subsection{
 SIS Gain Variations \& Redshift Dependence on Gain
                                 }
                                 
 The SIS performance has gradually degraded due to the accumulated energetic particle 
 radiation damage, which is manifested by  
 an increase in residual dark-current distribution (RDD) and in signal charge transfer inefficiency (CTI).
 Although the RDD effect is negligible in 1-CCD mode 
 (Dotani 2000 \footnote{www.astro.isas.ac.jp/$\sim$dotani/rdd.html})
 the increase of 
 the CTI reduces the apparent X-ray energy and thus causes
 a systematic variation of gain with time 
 (Dotani et al.\footnote{heasarc.gsfc.nasa.gov/docs/asca/newsletters/sis\_calibration5.html},
 Rasmussen et al. \footnote{heasarc.gsfc.nasa.gov/docs/asca/newsletters/sis\_performance2.html}).
 This effect is different for different chips resulting in enhancements of interchip global
 gain variations. 
  
 Correction for the above mentioned effect have been carried out using the 
 Ni background fluorescence line originated from the kovar (iron-nickel-cobalt alloy),
 which covers the frame store region of the chips
 and also using observations of Cas A. To investigate the 
 time dependence of the average CTI, the center energy of the Ni line has been 
 recorded every 3 months from standard chips (S0C1 and S1C3) in 1 
 CCD mode (Dotani et al.\footnote{heasarc.gsfc.nasa.gov/docs/asca/newsletters/sis\_calibration5.html}). 
 The decrease of the line energy is well approximated by a linear function of time.
 Furthermore, the gain variation with time was observed to vary non-uniformly for different 
 spectral lines
 after CTI correction. This may be a consequence of the assumption that the CTI is independent 
 of X-ray energy as well as from scatter of the line energies among the pointing positions. 
 
 The event files analyzed in this work were corrected by the tool ASCALIN V0.9t, using
 the most recent gain calibration file sisph2pi\_110397.fits. The line center energies are different
 from chip to chip by at most 2\% (interchip gain variation). Since we are showing 
 results for SIS 0 \& 1 separately, we do not worry about global gain variation differences
 between these detectors. However, we need to estimate 
 the gain variation across each CCD (intrachip gain variation). We did
 so by analyzing the gain scatter across the detectors measured using Cas A data. 
 After the standard correction analysis is completed,
 the systematic errors on the FeK$\alpha$ line centers are reduced to 
 $\sim$ 0.29\% (870km/s) for S0C1 and $\sim$ 0.28\% (840 km/s) for S1C3.
 
 The values mentioned above are based on a more recent preliminary calibration analysis 
 performed by Dotani and kindly provided to us by K. Mukai
 \footnote{see www.astro.lsa.umich.edu/$\sim$rdupke/dotani.ps}. 
 Using the older calibration references cited in the previous 
 paragraphs\footnote{heasarc.gsfc.nasa.gov/docs/asca/newsletters/rad\_dam\_sis3.html}
 would give a $\sigma_{gain}$ of 0.22\% and 0.38\% for SIS 0 \& SIS 1, respectively. 
 This discrepancy does not change any of the results obtained in this paper. 
 However, the larger observed uncertainties associated with the determination of 
 best-fit redshifts in SIS 1 suggests that $\sigma_{gain}$ for SIS 1 should be 
 greater than that for SIS0.
 
 We tested the sensitivity of our observations to possible residual gain 
 variations across the SISs using Monte Carlo simulations. Supposing the redshift 
 to be constant for the two regions where we find maximum velocity differences (P4 \& P6),
 we generated fake spectra for both 
 spectrometers and compared the best-fit redshift differences.
 Then we calculated the probability to find the same redshift differences 
 (or greater than) that we observed in the real regions , for SIS 0 \& 1.  The procedure
 used is completely analogous to that used in testing the effects of residual gain variations 
 across the GISs
 in the velocity measurements for the Perseus cluster (Dupke \& Bregman 2000, 2001) and we refer 
 the reader to that paper for details. 
 
 We simulated 1000 SIS 0 \& 1 spectra corresponding to the real observation using the same 
 spectral model, parameters and responses correspondent to the real regions. The simulated spectra 
 were then grouped, 
 fitted and had a gain uncertainty added to their best-fit redshift values.
 We assumed that the gain variations follow a Gaussian distribution with a standard deviation 
 ($\sigma_{gain}$), 
 which is different for each spectrometer, and a zero mean. We adopted as 
 our 1-$\sigma$ gain variations ($\sigma_{gain}$) for the SIS 0 \& 1 
 the values described above, i.e., 0.29\% and 0.28\%, respectively. We then calculated the 
 probability of finding
 the same (or greater) redshift differences between the two simulated regions that we found in
 the real ones, i.e., $\Delta$z = 0.009; that probability is found to be $\la$ 0.002.
 
 In Figure 4 we plot the dependence of the chance probability (described in the previous paragraph)
 on the assumed $\sigma_{gain}$ of the spectrometers (for the purpose of illustration $\sigma_{gain}$ 
 for SIS 0 is assumed to be the same as that for SIS 1, which roughly represents the real case). 
 In Figure 5 we
 also indicate the value for $\sigma_{gain}$ estimated in the previous paragraphs ($\sim$0.28\%).
 In our real case we observed not only two but at least three regions 
 (more than three if SIS 0 data alone is considered) with 
 discrepant redshifts (P3, P4 \& P6). Therefore, the significance of the velocity gradient 
 found in Centaurus obtained by 
 the method described above ($99.8\%$) should be taken as a lower limit.

 \subsection{Spectral Fittings Without the FeK Line Complex
                                        }
   
The velocity measurements presented in the previous section were derived from spectral fittings 
of SIS 0 \& 1 encompassing the usable energy range of the spectrometers, i.e., from
0.8 to 10 keV, and are mainly driven by the FeK$_{\alpha}$ line complex. Since the 
Centaurus cluster has very high ($\sim$solar) metal abundances it 
allows for the possibility of reliably measuring velocities using other spectral lines, or
at least, excluding the FeK line complex energy range. When we fitted the spectra from 
SIS 0 \& 1 for different energy ranges we noticed a systematic gain discrepancy
between spectral fittings that included FeK line complex from those excluding it. 
Excluding energies corresponding to both FeK and FeL complexes in the spectral fittings 
produced best-fit velocities (determined from spectral lines of 
other individual elements) that were more consistent with those fits that included the FeL line 
complex than with those that included the FeK line complex.
When the FeK lines were included the measured redshifts were consistently lower. As mentioned 
previously, this effect is most 
likely due to the assumption that the CTI is independent of X-ray 
energy during the in-house calibration process. 

In order to check how the observed velocity distributions were affected by the  
effect mentioned in the previous paragraph we performed a series of redshift 
measurements excluding the FeK region ($<$6 keV). 
The spectral fitting characteristics were completely analogous to that used in the last section.
The resulting best-fit values for temperature, metal abundance and radial velocities 
are listed in Table 4 and the redshift distributions for SIS 0 \& 1 are shown in Figures 5a,b.
The best-fit temperatures found when the FeK line complex is excluded from the analysis are, 
in general, slightly higher than, but similar to, those obtained using the full useable energy range. 
 The azimuthal temperature gradient is still marginally seen.
 The measured temperatures in the center are 2.84$\pm$0.03 keV (SIS0) \& 2.77$^{+0.03}_{-0.07}$ keV (SIS1), 
 rising to 3.70$\pm$0.12 keV (SIS0) \& 4.11$\pm$0.16 keV (SIS0) towards the direction
 of Cen 45. The abundance measurements are not well constrained enough to detect a central metal
 enhancement and are consistent with a flat profile ($\sim$ 1.25$\pm$0.25 Solar) with a large dispersion.
 
It can be seen from Figures 5a,b that, although the best-fit redshifts are typically higher than 
those in Figures 2a,b, 
the behavior of the azimuthal distributions with and without the FeK complex energy range are very 
similar. A significant
($>$ 90\% confidence) velocity gradient is also found. The velocity distribution reaches 
a minimum around regions P3 \& P4 and a 
maximum at region P6 \& P7. The velocity gradient is found to be steeper for SIS 0. 
The corresponding maximum 
velocity differences are $\sim$ 6.3$\pm$1.2 $\times$ 10$^{3}$ km~s$^{-1}$ for SIS 0 
and $\sim$ 3.18$\pm$0.96 $\times$ 10$^{3}$ km~s$^{-1}$ for SIS 1 (1-$\sigma$ errors).

 \subsection{SIS Observation of June 1993 (P93)
                                        }

In the previous sections we have shown that the velocity gradient in the ICM of 
the Centaurus cluster is detected with high significance using SIS 0 \& 1 data including 
and excluding the FeK line complex. We 
have also shown that it is not due to gain variations within the CCDs. Since {\sl ASCA} has 
observed the Centaurus cluster several times, we can test the reliability 
of the velocity measurements by performing the same analysis for another pointing, with 
different observing characteristics 
(e.g. observing date, CCD clock mode, position on the CCD, etc). Out of the two archived 
pointings in the {\sl ASCA} 
archive that cover the whole central regions of Centaurus, we chose the one with larger 
spatial coverage and exposure time. We refer to this pointing as P93 and list its 
characteristics in Table 1.

P93, however, has three main disadvantages when compared with P95. Firstly, it has been taken in 
4-CCD mode. The energy resolution of the SISs in 2-4 CCD modes has degraded by a factor of four 
with time with respect to 1-CCD mode observations due to the increase of RDD\footnote{
heasarc.gsfc.nasa.gov/docs/asca/newsletters/sis\_calibration5.html}. 
RDD effects were already not negligible for 4-CCD mode observations at the time P93 was taken
\footnote{heasarc.gsfc.nasa.gov/docs/asca/newsletters/sis\_performance2.html}(Rasmussen et al. 1994).  
Secondly, the effective exposure time is relatively short, around four times smaller than for P95; 
Thirdly,
the cluster's central regions are mainly extended over two chips in each spectrometer 
(S0C1 \& S0C2 and S1C3 \& S1C0).
These two chips have different overall gain off-sets,
requiring a more elaborated correction for gain variations, thus increasing the uncertainties 
of the measured redshifts. 
The first two of the above mentioned problems cause, effectively, larger statistical uncertainties
when determining velocities. The last problem is more serious. If not taken into account, 
interchip gain difference can be much larger than the velocities one is trying to measure, 
thus being able to artificially create, erase, or amplify velocity gradients.  

We corrected for interchip gain variations as follows: since the cluster's center is 
located very near the separation between chips, we selected a region with 3$^\prime$ radius 
centered on the X-ray peak and analyzed the spectra of the two parts of that region that fell on 
each separate chip; 3$^\prime$ at Centaurus distance corresponds to $\sim$57 h$_{50}^{-1}$ kpc. 
At that distance from the cluster center we assume that any velocity 
gradient that we measure between the two semi-regions is purely due to interchip gain variations.
If there were true velocity differences detectable by the spectrometers in that region 
($\gtrsim$600 km~s$^{-1}$) the gas would be gravitationally unbound. By comparing
the velocity difference between these two regions we found the correction factors that should 
be applied to each CCD so that the gain difference between chips did not affect the velocity 
measurements. To join data from both spectrometers (to improve statistics) we
 corrected for the global gain differences for each pair of chips in each spectrometer. In order 
 to bring all redshift values from all four chips to a unique solution 
within the central region we determined the gain correction factors that should be
applied simultaneously to S0C1, S0C2, S1C3 \& S1C0 as 1.0\%, 1.2\%, 0.6\% \& 1.0\%, respectively. 

After applying the gain correction described above we selected extraction regions analogous
to those used for P95, i.e. with 3$^\prime$ radius at 5$^\prime$ from the center with a $\sim$
20$^\circ$ rotation to compensate for the roll angle difference between the two pointings.  
If one of these 
regions crossed the chip boundaries and therefore could be separated in two parts
we ignored the part that had the smaller number of counts (the improvement in photon statistics 
does not justify the uncertainties generated by the gain correction procedures). \footnote{Within 
the extraction regions considered in the analysis of the central regions (circles
of 3$^\prime$ radius 5$^\prime$ from the center) some photons fell in the top 
two chips (in detector coordinates) of each 
spectrometer (S0C0 \& S0C3 and S1C2 \& S1C1). However, these regions 
are small and the gain correction procedure described above would create 
much higher uncertainties than what we would gain statistically by including those regions. 
Therefore, we consider photons detected only within the two chips described in the previous 
paragraph.}

The results are shown if Figure 6 and listed in Table 5. The temperature and abundance 
distributions are consistent with, though less significant than, those obtained for P95. 
The azimuthal temperature gradient outside
the cooling flow region is clearly seen. The maximum temperature is reached towards the
Cen 45 group (T=3.49$\pm$0.25 keV), the temperature within the central 3$^\prime$ is 
2.54$^{+0.07}_{-0.04}$ keV
and the central abundance gradient is marginally visible. The large errors associated with velocity 
measurements in P93 are due to: 1) the effective exposure being $\sim$ 4 times smaller than
in P95; 2) the observations being taken in 4-CCD mode, which worsens the energy resolution 
($\ga$3.5\% at 6.0 keV); 
3) the correcting procedure for interchip gain variations involves assumptions that may 
be oversimplified (e.g. the non-existence of true gas velocity differences within 
a 3$^\prime$ radius.
Nevertheless, {\it a velocity gradient 
is also seen roughly in the same orientation as that detected in P95!} In particular,
P3 \& P4 have significantly ($>$ 68\% confidence) lower redshifts than P6. 
The velocity difference correspondent to regions P4 \& P6 is 
$\Delta$V$\sim(1.8\pm1.5)\times10^{3}km~s^{-1}$ (1-$\sigma$ errors), consistent with the velocity 
measurements from P95.

                                        \section{
Discussion
                                         }
Spatially resolved spectral analysis of {\sl ASCA} SIS 0 \& 1 data carried out in this work 
strongly indicates the presence of significant azimuthal temperature, metal abundance and 
velocity gradients in the ICM of the Centaurus cluster. 
The temperature and metal abundance profiles derived in this work indicate the presence 
of a very mild cooling flow and a central abundance enhancement, consistent with 
those found in previous works. 
Both temperature and abundance gradients are steeper (with different signs) towards the 
Southeast direction, 
where Cen 45 is located. The increase in temperature and  mild dilution of metal abundances
towards Cen 45 is consistent with the hypothesis that Cen 45 is in the process of merging 
with the main body of the Centaurus cluster (Cen 30).

The velocity distribution shows a significant gradient along a direction roughly 
{\it perpendicular} to the direction of Cen 45 subgroup. The amplitude of the velocity gradient 
(along the P4-P6 direction) can be as high as $\sim$ 25 h$_{50}$ km~s$^{-1}$kpc$^{-1}$. 
The velocity gradient is 
significantly ($>$90\% confidence) detected with both SISs on-board {\sl ASCA}. The velocity 
gradient is also 
seen in a different {\sl ASCA} observation performed $\sim$ 2 years earlier, although with smaller
significance. We have shown in the previous sections that the velocity gradient is not
caused by gain fluctuations across the SIS chips, thus suggesting the presence of 
large-scale bulk motions of the intracluster gas in this cluster. 

A tentative global velocity distribution for the central regions of the Centaurus cluster that 
summarizes the results obtained in this paper is plotted in Figure 7 and listed in Table 6. 
The data points show a weighted average velocity departure from the median $\Delta$V for all 
cases that we have analyzed previously, i.e., SIS 0 \& 1 for P95, SIS 0 \& 1 
for P95 without the FeK line complex and SIS 0 \& 1 for P93. The errors include the fitting errors and 
a constant gain uncertainty corresponding to the 1-$\sigma$ 
intrachip gain fluctuation described in the previous sections, i.e., 0.28\%. 
The direction towards Cen 45 is shown by the vertical lines and the uncertainties 
of the velocity measurements for the very central regions is indicated by the horizontal lines. 
For illustration, we fit a simple solid body rotation function to the data and obtain a 
best-fit maximum rotational velocity of 1.59$\pm$0.32 $\times$ 10$^{3}$ km~s$^{-1}$.
\footnote{It should be noted that bulk velocities as high as these can be, in principle, measured also
through high resolution differential measurements of the Sunyaev-Zel'dovich effect across the line 
of maximum velocity difference, especially in cool clusters such as Centaurus, where the thermal component 
is weakly dominant below 300 GHz and $(\frac{\Delta T}{T})_{CMBR} \sim 10^{-4}-10^{-3}$.}

Similarly to the case of the Perseus cluster, the large scale X-ray 
elongation is near perpendicular to the direction where the velocity gradient is maximum.
However, in the case of Centaurus, the elongation towards Cen 45 can be directly associated
with the elongation of the gravitational well along the collision axis. At smaller scales
($\la$ 10$^\prime$) there is evidence for isophotal twisting (Mohr, Fabricant \& Geller 1993) 
and the major axis of the surface brightness elongation shifts $\sim$ 50$^\circ$ towards the 
direction of maximum gas velocity gradient.

One way of generating such large angular momentum in clusters is through 
cluster-cluster mergers. Several recent numerical simulations of 
off-center cluster mergers predict residual 
intracluster gas bulk rotation with velocities of a few thousand km~s$^{-1}$ 
(Ricker 1998; Roettiger, Stone, \& Mushotzky 1998; 
Takizawa 1999, 2000; Roettiger \& Flores 2000) indicating that 
a high fraction of the merging energy can be transferred to gas rotation 
(Pearce, Thomas, \& Couchman, 1994). In the case of Centaurus there is growing
evidence indicating that this cluster is in the process
of merging with the galaxy group Cen 45 including:
(1) the galaxy radial velocity bimodality detected by Lucey et al. (1986a,b), (2) the large scale gas radial 
velocity bimodality shown in this work (GIS analysis);
(3) the large scale temperature gradient detected towards NGC 4709, which is
expected from heating due to interaction between the two subgroups (Churazov et al. 1999);
(4) the small scale azimuthal temperature and abundance gradients detected towards Cen 45 (this work);
(5) the morphological segregation both in velocity and physical space: 
Cen 30 is dominated by early type galaxies and Cen 45 by late types (Stein, Jerjen \& Federspiel 1997);

Simulations of head-on mergers show that the core of the primary 
cluster changes from being spherical before core-crossing to an elongated shape along 
the merger direction after 
core-crossing (Gomez 2000). This is not observed in the core of Cen 30 
(the X-ray elongation in the core is actually near perpendicular to the collision axis).
In a head-on collision with a smaller 
mass sub-unit, the core of the main cluster may display cooler regions towards the side opposite to the
strongest bow shock (at the boundary of the two systems) shortly (0.25 Gyr) after core crossing 
(e.g. Gomez 2000). This would be consistent with the smooth azimuthal temperature gradient along the 
direction towards Cen 45 (P1, P0, P5). However, this scenario, i.e., 
we are in a post-core crossing epoch, faces several difficulties. 
Firstly, this is a very short period of time in order to establish full rotation. 
Secondly, even if rotation is not fully established significant 
gas bulk motions are seen only after the boundary shock has encompassed the cores of the two
sub-systems (Ricker 1998), 
and thirdly, galaxies in the two sub-systems in Centaurus are still in a clearly segregated state.

Even though it has been suggested that mergers erase cooling flows 
(e.g. Edge, Steward \& Fabian 1992, Roettiger et al. 1993) and central abundance gradients, 
recent simulations of cooling-flow-cluster mergers show that, even in head-on mergers, 
the bow shock that appears in between the colliding systems 
can protect the gas from mixing and cooling flows can survive mergers depending 
on the produced ram-pressure of the gas in 
the infalling cluster (Gomez et al. 2000, see also Fabian \& Daines 1991). Cooling flow survival in 
mergers is predicted to be even more probable in {\it off-center} mergers 
(Gomez et al. 2000). Further, large scale merger-induced gas rotation can last for a few 
crossing times (Roettiger \& Flores 2000), which suggests a more likely scenario 
to explain the gas velocity distribution in Centaurus, that is, the velocity 
gradient in Centaurus pre-existed 
the current merger and is not connected to the recent infall of Cen 45. In this scenario the
observed velocity gradient was generated by some previous off-center merger, that happened 
$>$ 0.2-0.6 h$^{-1}_{50}$ Gyr (circulation time) ago and did not 
destroy either the cooling flow or the central abundance gradient\footnote{After this paper was submitted
we learned that an extensive gas temperature analysis of multiple pointings of the Centaurus cluster 
by Furusho et al. (2001) also supports this scenario.}.  This past merger event 
could have been a reasonably 
isolated event, in which case the temperature asymmetry in the maps of Churazov et al.(1999) could 
actually be due to the effects of a current head-on collision with a previously rotating cluster. 
If this is the case, Centaurus may provide 
a crucial observational basis for future comparison to (currently unavailable) numerical simulations of 
cluster mergers where the components have a previously imparted intrinsic angular momentum.

However, the velocity structure observed could be also created by a past series of matter accretion 
events from filamentary structures, in the intersection of which clusters are believed to form. The poor 
spatial resolution of the spectrometers on-board {\sl ASCA} does not allow us to construct a 
detailed velocity 
map that would discriminate between a more or less systematic ICM ``rotation'' from a 
more chaotic ICM ``stormy weather'' as predicted by Burns et al. 1999. 
Multiple off-center long exposure velocity measurements of intracluster gas with the {\sl Chandra} 
and {\sl XMM-Newton} satellites will allow us to reduce the systematic uncertainties related to 
gain variations, and consequently, 
to determine ICM velocities in the central and outer regions more precisely. This, combined 
with more specific cluster-cluster merger simulations will allow us to strongly constrain 
the current evolutionary stage of the Centaurus cluster.

                           \section{
Summary
}
We have used, for the first time, spatially resolved spectroscopy for determination 
of ICM radial-velocities in the Centaurus cluster. The main results of 
our velocity analysis are:
\begin{enumerate}
\item The gas velocity distribution at $>$ 8$^{\prime}$ determined with the GISs 
shows that the region associated with Cen 45 has a radial velocity higher than 
that of the rest of the cluster by $\sim$ 1.77$\pm$0.96 $\times$ 10$^{3}$ km~s$^{-1}$.
This is consistent with the optically determined velocity differences between Cen 45 and Cen 30
of $\sim$ 1.35$\pm$0.19 $\times$ 10$^{3}$ km~s$^{-1}$ (Stein et al. 1997).
\item We found azimuthal temperature and abundance gradients in the cluster's central  
regions ($<$ 8$^{\prime}$). The azimuthal temperature variations are in agreement with the 
temperature maps determined previously by other authors (Churazov et al. 1999). Both temperature
and metal abundance variations seem to be directly related to the infall of the Cen 45 subgroup. 
\item The gas velocity distribution at $<$ 8$^{\prime}$ determined with the SISs reveals 
a significant velocity gradient ($>$99.9\% confidence) along a direction near-perpendicular
to the direction towards Cen45. This detection is found to be significant in both SIS 0 \& 1 separately
 and also in another {\sl ASCA} pointing taken two years earlier.
 The velocity gradient does not show any clear correlation
with the temperature or metal abundance gradients. It is not due to spectral modeling effects
or to intrachip gain fluctuations. The velocity gradient is consistent with gas bulk rotation 
with circular velocity 1.59$\pm$0.32 $\times$ 10$^{3}$ km~s$^{-1}$, 
implying that gas kinetic energy is comparable to the thermal energy in the intracluster gas of Centaurus.
\item Comparison to cluster-merger numerical simulations, available in the literature, 
suggests that the velocity gradients that we detected in the central regions of 
Centaurus are not associated with the current infall of the Cen 45 subgroup. Therefore, we 
believe that it is more likely that the central velocity gradient is remnant of one or more previous 
off-center merger events.
\end{enumerate}

\acknowledgments We would like to thank J. Irwin for the many helpful discussions 
and suggestions and the anonymous referee for useful suggestions. We particularly thank
K. Mukai for providing information about {\sl ASCA} SIS gain calibrations that 
was crucial to this work. We acknowledge support 
from NASA Grant NAG 5-3247. This research made use of the HEASARC 
{\sl ASCA} database and NED.

                                
                                \clearpage
                                \begin{figure}
                                \title{
Figure Captions
                                }
\caption{
Distribution of the spatial regions for the pointings analyzed in this work. Regions labeled with a P$_{i}$, where i=0--8,
denote the set of extraction regions used in the SIS 0 \& 1 analysis of the central regions for P95 and P93. The radius of each 
circular region is 3$\arcmin$ and they are 5$\arcmin$ away from the cluster's center. The larger regions denoted J12, J34,
J56 and J78 are combinations two-by-two of circular regions 12$\arcmin$ away with 5$\arcmin$ radius used for the GIS velocity analysis
of the outer regions of Centaurus. For the same analysis a circular region 11$\arcmin$ away with 8$\arcmin$ radius
was also selected and is denoted by C45. The approximate position of NGC 4709 is towards the 
bottom left of C45. Surface brightness contours of Centaurus are overlaid.
                                }
\caption{
a)Azimuthal distribution of temperature (TOP), metal abundance (MIDDLE), and redshift (BOTTOM) as 
a function of the position angle (radians) obtained from the analysis of SIS 0 data for
P95. The first data point from the left for all plots corresponds
to P1, increasing to P8 (last). For all plots dark solid horizontal lines represent the 
90\% confidence limits for the central region (P0). The region between vertical lines 
indicates the approximate direction of Cen 45. Errors for all plots are 90\% confidence.
A velocity difference equivalent to 2700 km~s$^{-1}$ is also indicated by horizontal dashed lines. 
b)Same as (a) but for SIS 1
                                }
\caption{
Confidence contour plot for two interesting parameters (redshifts) for regions P4 \& P6 using 
SIS 0 \& 1 data for P95. The two contours 
correspond to 90\% and 99\% confidence levels. The line of equal redshifts 
is also indicated. The contours are found from simultaneous spectral fittings of four data groups 
(P4 SIS 0\&1 and P6 SIS 0\&1). The redshifts parameters for both instruments are tied together for 
the spectral fittings of the same region.
                                }
                               \caption{
Probability of finding redshift differences equal or greater than what we 
observed for the real data (P4$\times$P6) by chance as a function of the standard deviation of intrachip gain 
variations for the SISs. The $\sigma_{gain}$ for SIS 0 is assumed to be the same as that for SIS 1
in this plot for illustration purposes. The actual value of $\sigma_{gain}$ used in this paper ($\sim$0.28\%) is also 
indicated.
                                }
\caption{
a)Azimuthal distribution of redshifts as a function of the position angle (radians) obtained from the 
spectral fittings of SIS 0 data for P95 when the FeK line complex energy range was excluded. 
First data point from the left for all plots corresponds
to P1, increasing to P8 (last). For all plots dark solid horizontal lines represent the 
90\% confidence limits for the central region (P0). The region between vertical lines 
indicate the approximate direction of Cen 45. Errors for all plots are 90\% confidence. 
A velocity difference equivalent to 6300 km~s$^{-1}$ is also indicated by horizontal dashed lines.
b)Same as (a) but for SIS 1 and the horizontal dashed lines indicated a velocity difference 
equivalent to 4300 km~s$^{-1}$.   
                                }

\caption{
Azimuthal distribution of Temperature (TOP), Metal Abundance (MIDDLE), and Redshift (BOTTOM) as 
a function of the position angle (radians) obtained from the simultaneous spectral fittings of SIS 0 \& 1 
data for P93. The first data point from the left for all plots corresponds
to P1, increasing to P8 (last). The errorbars indicated show 90\% confidence errors for 
temperature and abundances and 68\%(1$\sigma$) for redshifts. The dark solid horizontal 
lines represent the confidence limits for the central region (P0). The region between vertical lines 
indicate the approximate direction of Cen 45. A velocity difference equivalent to 1800 km~s$^{-1}$ 
is also indicated by horizontal dashed lines.
                                }
\caption{
Average azimuthal gas radial velocity distribution for the central regions of the Centaurus cluster. Data points indicate 
the weighted average of the departure from the median. 
The errors also include a 1$\sigma$ gain variation of 0.28\% (typical for the SIS chips at 1995) for all cases 
(all instruments, all pointings) analyzed in this work. 
The first data point from the left for all plots corresponds to P1, increasing to P8 (last). 
The dark solid horizontal lines represent the 68\% confidence limits
for the central region (P0). The region between the two vertical lines indicates the approximate direction of Cen 45. 
The solid line indicate the bestfit for a simple solid-body cosine function with a corresponding maximum rotational 
velocity of $\sim$ 1600$\pm$300km s$^{-1}$.
                                }
\end{figure}
\clearpage

\begin{deluxetable}{lccccccc}
\small
\tablewidth{0pt}
\tablecaption{X-ray Observations}
\tablehead{
\colhead{Pointing} &
\colhead{Sequence}  &
\colhead{Date}  &
\colhead{RA} &
\colhead{DEC} &
\colhead{GIS EXP\tablenotemark{a}} &
\colhead{SIS EXP\tablenotemark{b}}  \nl
\colhead{} &
\colhead{Number} &
\colhead{Observed}  &
\colhead{(2000)} &
\colhead{(2000)} &
\colhead{(ksec)} &
\colhead{(ksec)} &
}
\startdata
P95 & $83026000$ & 1995-07-19 & 12h48m21.60s& -41$^{\circ}$19$\arcmin$44.8$\arcsec$ & 56.4 & 67.9 \nl
P93 & $80032000$ & 1993-06-30 & 12h48m24.00s& -41$^{\circ}$17$\arcmin$16.4$\arcsec$ & 14.7 & 16.6 \nl

\enddata
\tablenotetext{a}{Effective Exposure (Average for GIS 2 \& 3)}
\tablenotetext{b}{Effective Exposure (Average for SIS 0 \& 1)}
\end{deluxetable}

\begin{deluxetable}{lccccc}
\small
\tablewidth{0pt}
\tablecaption{GIS Spectral Fittings for P95\tablenotemark{a,b}}
\tablehead{
\colhead{Region} &
\colhead{Temperature\tablenotemark{c}}  &
\colhead{Abund } &
\colhead{Redshift } &
\colhead{$\chi^{2}_{\nu}$} & \nl
\colhead{} &
\colhead{(keV)} &
\colhead{(Solar)} &
\colhead{(10$^{-2}$)} &
\colhead{} &
}
\startdata
C45      & 5.47$^{+0.28}_{-0.27}$  & 0.50$^{+0.09}_{-0.09}$  &  3.27$^{+1.58}_{-1.13}$ & 1.062\nl
All-C45  & 3.44$^{+0.03}_{-0.03}$  & 0.77$^{+0.03}_{-0.03}$  &  1.81$^{+0.05}_{-0.01}$ & 1.206\nl
J0       & 2.73$^{+0.02}_{-0.02}$  & 1.09$^{+0.05}_{-0.05}$  &  1.81$^{+0.05}_{-0.01}$ & 1.102\nl
J12      & 3.67$^{+0.09}_{-0.08}$  & 0.54$^{+0.07}_{-0.06}$  &  2.35$^{+0.37}_{-0.52}$ & 0.983\nl
J34      & 4.16$^{+0.13}_{-0.13}$  & 0.47$^{+0.08}_{-0.08}$  &  1.93$^{+1.07}_{-0.08}$ & 0.969\nl
J56      & 4.67$^{+0.17}_{-0.17}$  & 0.44$^{+0.09}_{-0.08}$  &  2.40$^{+0.60}_{-0.50}$ & 0.969\nl
J78      & 3.83$^{+0.10}_{-0.09}$  & 0.42$^{+0.06}_{-0.06}$  &  1.61$^{+0.26}_{-0.65}$ & 0.970\nl
\enddata
\tablenotetext{a}{Errors are 90\% confidence}
\tablenotetext{b}{0.8--10.0 keV}
\tablenotetext{c}{N$_{H}$ fixed at the Galactic value}
\end{deluxetable}

\begin{deluxetable}{lccccccc}
\small
\tablewidth{0pt}
\tablecaption{Spectral Fittings for P95\tablenotemark{a,b,c}}
\tablehead{
\colhead{Region} &
\colhead{T(SIS0)\tablenotemark{d,e}}  &
\colhead{T(SIS1)\tablenotemark{d}}  &
\colhead{Abund (SIS0)} &
\colhead{Abund (SIS1)} &
\colhead{z (SIS0)} &
\colhead{z (SIS1)}  \nl
\colhead{} &
\colhead{(keV)} &
\colhead{(keV)} &
\colhead{(Solar)} &
\colhead{(Solar)} &
\colhead{(10$^{-2}$)} &
\colhead{(10$^{-2}$)} &
}
\startdata
P0  & 2.67$^{+0.03}_{-0.03}$ & 2.69$^{+0.06}_{-0.05}$ & 1.13$^{+0.06}_{-0.03}$ & 1.23$^{+0.07}_{-0.06}$ & 1.32$^{+0.02}_{-0.01}$ & 1.41$^{+0.04}_{-0.02}$\nl
P1  & 2.94$^{+0.06}_{-0.06}$ & 2.74$^{+0.08}_{-0.06}$ & 1.11$^{+0.12}_{-0.10}$ & 1.10$^{+0.15}_{-0.12}$ & 1.33$^{+0.15}_{-0.10}$ & 1.58$^{+0.38}_{-0.23}$\nl
P2  & 3.11$^{+0.07}_{-0.06}$ & 3.02$^{+0.07}_{-0.07}$ & 0.96$^{+0.09}_{-0.09}$ & 1.11$^{+0.12}_{-0.11}$ & 1.33$^{+0.16}_{-0.11}$ & 1.40$^{+0.08}_{-0.05}$ \nl
P3  & 3.43$^{+0.09}_{-0.08}$ & 3.54$^{+0.09}_{-0.05}$ & 0.83$^{+0.08}_{-0.08}$ & 1.04$^{+0.11}_{-0.10}$ & 0.88$^{+0.15}_{-0.05}$ & 0.88$^{+0.15}_{-0.05}$  \nl
P4  & 3.59$^{+0.08}_{-0.09}$ & 3.84$^{+0.11}_{-0.12}$ & 0.89$^{+0.09}_{-0.08}$ & 0.84$^{+0.11}_{-0.10}$ & 0.88$^{+0.19}_{-0.05}$ & 0.85$^{+0.34}_{-0.12}$ \nl
P5  & 3.56$^{+0.10}_{-0.09}$ & 3.86$^{+0.13}_{-0.12}$ & 0.83$^{+0.09}_{-0.10}$ & 0.87$^{+0.13}_{-0.12}$ & 1.33$^{+0.12}_{-0.09}$ & 1.33$^{+0.16}_{-0.51}$ \nl
P6  & 3.36$^{+0.09}_{-0.09}$ & 3.37$^{+0.11}_{-0.11}$ & 0.86$^{+0.10}_{-0.09}$ & 0.99$^{+0.14}_{-0.13}$ & 1.79$^{+0.10}_{-0.04}$ & 1.76$^{+0.26}_{-0.38}$\nl
P7  & 2.96$^{+0.08}_{-0.08}$ & 2.96$^{+0.10}_{-0.09}$ & 0.94$^{+0.05}_{-0.12}$ & 0.93$^{+0.13}_{-0.12}$ & 1.78$^{+0.10}_{-0.06}$ & 0.87$^{+0.90}_{-0.03}$\nl
P8  & 2.93$^{+0.07}_{-0.06}$ & 2.88$^{+0.08}_{-0.07}$ & 1.02$^{+0.14}_{-0.06}$ & 1.01$^{+0.12}_{-0.11}$ & 1.78$^{+0.07}_{-0.04}$ & 1.40$^{+0.25}_{-0.08}$\nl

\enddata
\tablenotetext{a}{Errors are 90\% confidence level}
\tablenotetext{b}{0.8--10.0 keV}
\tablenotetext{c}{$\chi^{2}_{\nu}\sim1.1$}
\tablenotetext{d}{N$_{H}$ fixed at the Galactic value}
\tablenotetext{e}{Global redshifts boosted up by 0.9\%}
\end{deluxetable}

\begin{deluxetable}{lccccccc}
\small
\tablewidth{0pt}
\tablecaption{Spectral Fittings for P95\tablenotemark{a,b,c}}
\tablehead{
\colhead{Region} &
\colhead{T(SIS0)\tablenotemark{d,e}}  &
\colhead{T(SIS1)\tablenotemark{d}}  &
\colhead{Abund (SIS0)} &
\colhead{Abund (SIS1)} &
\colhead{z (SIS0)} &
\colhead{z (SIS1)}  \nl
\colhead{} &
\colhead{(keV)} &
\colhead{(keV)} &
\colhead{(Solar)} &
\colhead{(Solar)} &
\colhead{(10$^{-2}$)} &
\colhead{(10$^{-2}$)} &
}
\startdata
P0$_{FeL}$  & 2.84$^{+0.03}_{-0.03}$ & 2.77$^{+0.03}_{-0.07}$ & 1.46$^{+0.06}_{-0.06}$ & 1.43$^{+0.07}_{-0.08}$ & 2.71$^{+0.02}_{-0.00}$ & 2.37$^{+0.04}_{-0.07}$\nl
P1$_{FeL}$  & 3.09$^{+0.09}_{-0.09}$ & 2.86$^{+0.05}_{-0.11}$ & 1.40$^{+0.15}_{-0.16}$ & 1.34$^{+0.25}_{-0.20}$ & 2.23$^{+0.47}_{-0.19}$ & 2.35$^{+0.07}_{-0.09}$\nl
P2$_{FeL}$  & 3.24$^{+0.05}_{-0.10}$ & 3.18$^{+0.10}_{-0.09}$ & 1.19$^{+0.17}_{-0.14}$ & 1.38$^{+0.20}_{-0.18}$ & 1.63$^{+0.69}_{-0.08}$ & 2.36$^{+0.06}_{-0.06}$ \nl
P3$_{FeL}$  & 3.53$^{+0.11}_{-0.12}$ & 3.82$^{+0.07}_{-0.13}$ & 1.00$^{+0.17}_{-0.13}$ & 1.55$^{+0.07}_{-0.23}$ & 1.19$^{+0.67}_{-0.07}$ & 2.32$^{+0.09}_{-0.55}$  \nl
P4$_{FeL}$  & 3.70$^{+0.13}_{-0.11}$ & 4.11$^{+0.17}_{-0.16}$ & 1.13$^{+0.16}_{-0.15}$ & 1.23$^{+0.28}_{-0.18}$ & 1.71$^{+0.14}_{-0.19}$ & 1.33$^{+0.34}_{-0.50}$ \nl
P5$_{FeL}$  & 3.71$^{+0.14}_{-0.14}$ & 3.99$^{+0.19}_{-0.18}$ & 1.04$^{+0.20}_{-0.16}$ & 1.09$^{+0.14}_{-0.21}$ & 2.57$^{+0.82}_{-0.40}$ & 1.37$^{+0.91}_{-0.14}$ \nl
P6$_{FeL}$  & 3.59$^{+0.10}_{-0.10}$ & 3.55$^{+0.12}_{-0.19}$ & 1.29$^{+0.23}_{-0.23}$ & 1.25$^{+0.25}_{-0.26}$ & 3.60$^{+0.47}_{-0.17}$ & 2.72$^{+0.18}_{-0.98}$\nl
P7$_{FeL}$  & 3.11$^{+0.14}_{-0.12}$ & 3.23$^{+0.15}_{-0.15}$ & 1.17$^{+0.24}_{-0.17}$ & 1.34$^{+0.29}_{-0.24}$ & 3.07$^{+0.54}_{-0.51}$ & 2.54$^{+0.63}_{-0.44}$\nl
P8$_{FeL}$  & 3.13$^{+0.07}_{-0.09}$ & 3.07$^{+0.13}_{-0.11}$ & 1.45$^{+0.19}_{-0.10}$ & 1.33$^{+0.27}_{-0.18}$ & 2.63$^{+0.22}_{-0.08}$ & 2.28$^{+0.09}_{-0.48}$\nl

\enddata
\tablenotetext{a}{Errors are 90\% confidence level}
\tablenotetext{b}{FeK complex excluded (0.8-6 keV)}
\tablenotetext{c}{$\chi^{2}_{\nu}\sim1.1$}
\tablenotetext{d}{N$_{H}$ fixed at the corresponding Galactic value}
\tablenotetext{e}{Global redshifts boosted up by 1.2\%}
\end{deluxetable}

\begin{deluxetable}{lccccc}
\small
\tablewidth{0pt}
\tablecaption{Spectral Fittings for P93\tablenotemark{a,d}}
\tablehead{
\colhead{Region} &
\colhead{Temp\tablenotemark{c}}  &
\colhead{Abund \tablenotemark{c}} &
\colhead{z\tablenotemark{b}} &
\colhead{$\chi^{2}_{\nu}$} & \nl
\colhead{} &
\colhead{(keV)} &
\colhead{(Solar)} &
\colhead{(10$^{-2}$)} &
\colhead{} &
}
\startdata
P0  & 2.54$^{+0.07}_{-0.04}$ & 1.02$^{+0.05}_{-0.08}$ & 1.02$^{+0.02}_{-0.02}$ & 1.19\nl
P1  & 3.00$^{+0.17}_{-0.14}$ & 0.98$^{+0.17}_{-0.17}$ & 1.09$^{+0.15}_{-0.15}$ & 1.125\nl
P2  & 3.30$^{+0.14}_{-0.15}$ & 0.91$^{+0.13}_{-0.11}$ & 1.15$^{+0.44}_{-0.06}$ & 1.132\nl
P3  & 3.35$^{+0.16}_{-0.15}$ & 0.81$^{+0.12}_{-0.12}$ & 1.00$^{+0.03}_{-0.07}$ & 0.994\nl
P4  & 3.46$^{+0.20}_{-0.20}$ & 0.87$^{+0.17}_{-0.12}$ & 0.75$^{+0.27}_{-0.28}$ & 1.095\nl
P5  & 3.49$^{+0.26}_{-0.24}$ & 0.75$^{+0.16}_{-0.17}$ & 0.77$^{+0.79}_{-0.21}$ & 1.013\nl
P6  & 3.27$^{+0.17}_{-0.16}$ & 0.71$^{+0.12}_{-0.09}$ & 1.32$^{+0.27}_{-0.22}$ & 0.995\nl
P7  & 3.13$^{+0.13}_{-0.13}$ & 0.74$^{+0.10}_{-0.10}$ & 1.11$^{+0.31}_{-0.15}$ & 1.055\nl
P8  & 2.80$^{+0.20}_{-0.16}$ & 0.85$^{+0.18}_{-0.14}$ & 1.26$^{+0.76}_{-0.06}$ & 1.399\nl
\enddata
\tablenotetext{a}{N$_{H}$ fixed at the Galactic value}
\tablenotetext{b}{Errors are 68\% confidence level}
\tablenotetext{c}{Errors are 90\% confidence level}
\tablenotetext{d}{Slope boosting correction for interchip gain differences $\left\{\begin{array}{cc}S0C1 & S0C2\\ S1C3 & S1C0\\ \end{array} \right\}=\left\{\begin{array}{cc} 1.01 & 1.012\\ 1.006 & 1.01 \\ \end{array} \right\}$}
\end{deluxetable}

\begin{deluxetable}{lcc}
\small
\tablewidth{0pt}
\tablecaption{Weighted Average Velocity Differences for All Pointings \tablenotemark{a}}
\tablehead{
\colhead{Region} &
\colhead{$\Delta$V}  & \nl
\colhead{} &
\colhead{(10$^{3}$~km~s$^{-1}$)} &
}
\startdata

P0  &  0.07$\pm$0.45\nl
P1  & -0.06$\pm$0.52\nl
P2  & -0.63$\pm$0.47\nl
P3  & -1.84$\pm$0.45\nl
P4  & -2.04$\pm$0.54\nl
P5  & -0.36$\pm$0.74\nl
P6  &  1.38$\pm$0.53\nl
P7  &  0.71$\pm$0.61\nl
P8  &  0.46$\pm$0.49\nl
\enddata
\tablenotetext{a}{1$\sigma$ errors include spectral fitting as well as intrachip gain variations.}
\end{deluxetable}

                               
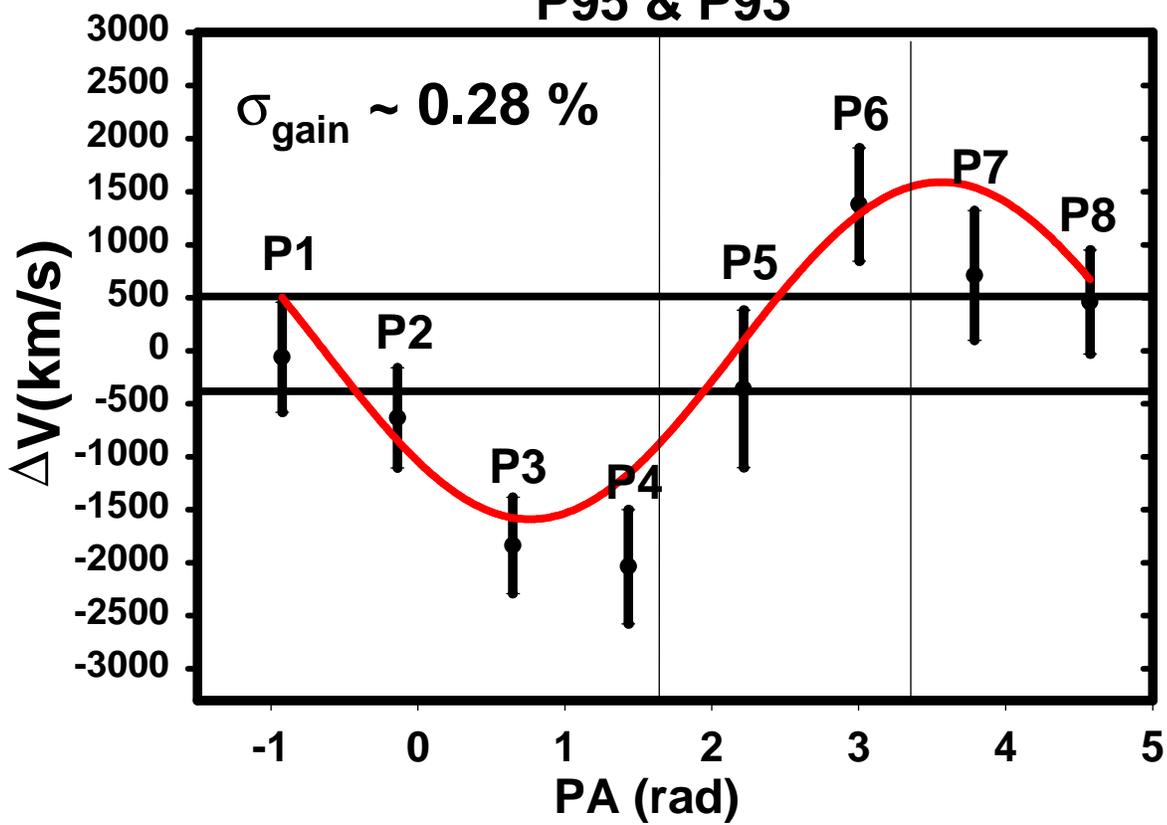
\end{document}